\documentclass[letterpaper,twocolumn,english,PRX]{revtex4-2}
\usepackage[LGR,T1]{fontenc}
\usepackage[utf8]{inputenc}
\usepackage{textcomp}
\usepackage{amstext}
\usepackage{amssymb}
\usepackage{graphicx}
\usepackage{wasysym}

\makeatletter

\pdfpageheight\paperheight
\pdfpagewidth\paperwidth

\ProvideTextCommand{\~}{LGR}[1]{\char126#1}

\newcommand{\lyxmathsym}[1]{\ifmmode\begingroup\def\b@ld{bold}
  \text{\ifx\math@version\b@ld\bfseries\fi#1}\endgroup\else#1\fi}

\mathchardef\mhyphen="2D

\makeatother

\usepackage{babel}
\begin{document}
\title{Synchronized oscillations in swarms of nematode Turbatrix aceti}
\author{Anton Peshkov}
\email{apeshkov@ur.rochester.edu}

\affiliation{Department of Physics and Astronomy, University of Rochester, Rochester,
NY 14627, USA}
\author{Sonia McGaffigan}
\affiliation{Department of Physics and Astronomy, University of Rochester, Rochester,
NY 14627, USA}
\author{Alice C. Quillen}
\affiliation{Department of Physics and Astronomy, University of Rochester, Rochester,
NY 14627, USA}
\begin{abstract}
There is a recent surge of interest in the behavior of active particles
that can at the same time align their direction of movement and synchronize
their oscillations, known as \emph{swarmalators}. While theoretical
and numerical models of such systems are now abundant, no real-life
examples have been shown to date. We present an experimental investigation
of the collective motion of the nematode \emph{Turbatrix aceti} that
self-propel by body undulation. We discover that these nematodes can
synchronize their body oscillations, forming striking traveling metachronal
waves, which produces strong fluid flows. We uncover that the location
and strength of this collective state can be controlled through the
shape of the confining structure; in our case the contact angle of
a droplet. This opens a way for producing controlled work such as
on-demand flows or displacement of objects. We illustrate this by
showing that the force generated by this state is sufficient to change
the physics of evaporation of fluid droplets, by counteracting the
surface-tension force, which allow us to estimate its strength. The
relatively large size and ease of culture make \emph{Turbatrix aceti}
a promising model organism for experimental investigation of swarming
and oscillating active matter capable of producing controllable work.
\end{abstract}
\maketitle

\section*{Introduction}

Collections of biological organisms can be considered active materials
\citep{Marchetti_2013} as energy is continuously dispersed through
their motion. Two kinds of collective behavior can be distinguished
for such organisms. On one hand, the self-propulsion of the organisms
can lead to collectively moving states such as ``turbulence'' in
bacterial suspensions \citep{SOKOLOV:2009:ID560}, flocking of birds
\citep{BALLERINI:2008:ID610} or schools of fishes \citep{Calovi_2014}.
On the other hand, some organisms performing periodic actions can
synchronize their oscillations, such as the synchronous flashing of
bugs \citep{Buck_1966}, crowd synchrony of pedestrians walking on
a bridge \citep{Strogatz_2005} or flagella of microorganisms that
beat in phase with one another \citep{Taylor_1951}. The latter example
is particularly interesting as it can lead not only to ``in-phase''
synchronization but also to ``moving phase'' or traveling motion
known as metachronal waves \citep{Niedermayer_2008,Brumley_2012,Elgeti_2013}.

A model that combines these two types of collective motion into particles
that at the same time self-propel and synchronize their oscillations
have recently been proposed in the form of swarmalators studied in
\citep{OKeeffe_2017}. In this model, the position of the particles
and the state of the oscillator are interdependent, which distinguish
it from previous studies where these two quantities were assumed to
be independent. This model has since been complemented by more analytical
and numerical investigations \citep{KRUK:2018:ID968,LEVIS:2019:ID969,JIMENEZ-MORALES:2020:ID971,KRUK:2020:ID970},
which found states with locally synchronized oscillations and particles
traveling in waves. However, none of these studied models have shown
a traveling wave state\emph{.} A possible experimental realization
of swarmalators\emph{ }has been proposed and numerically studied in
\citep{MANNA:2021:ID967} in the form of undulating synthetic active
flexible sheets, reminiscent of our nematodes. However, no experimental
realization of collectively moving and oscillating particles has been
demonstrated until this time.

In this study we report on collective behavior in a system of undulating
nematodes \emph{Turbatrix aceti (T. aceti)} commonly known as vinegar
eels. The vinegar eels are widely used in aquaculture as food for
young fishes and crustaceans. Therefore, they can be easily sourced
from aquarium supplies stores and their culture methods are straightforward.
The nematodes need to undulate to self-propel, and as we show in this
paper, the synchronization of these oscillations leads to the formation
of a collective metachronal wave\emph{.} While these waves are similar
to the one observed in cilia, the vinegar eels are not affixed to
the wall, and can exit and enter the wave which slowly moves along
the border.

The collective behavior of mobile particles can be sensitive to the
number of spatial dimensions in which they evolve \citep{Nguyen_2012,Chuang_2016},
as well as to confinement and container geometry due to interactions
with a boundary \citep{Hernandez2005,BRICARD:2015:ID680,Lushi_2014,Quillen_2020}.
We show that the formation of the metachronal wave by \emph{T. aceti}
in a three-dimensional space is induced by confinement, in our case
that of a contact angle of the droplet in which they swim. The the
confinement geometry not only controls the location of the collective
state, but also the number of nematodes involved in collective beating
and therefore the strength of the produced flow.

Past experimental investigation using bacterial suspensions have been
able to displace objects\citep{SOKOLOV:2010:ID620,DI:2010:ID883},
extract energy\citep{VIZSNYICZAI:2017:ID885} or produce controlled
organized flows\citep{GAO:2015:ID886}. However, extraction of energy
and flow productions has only been achieved by trapping individual
bacteria in special cells and using their outside sticking flagella
as motors \citep{VIZSNYICZAI:2017:ID885,GAO:2015:ID886}. While cilia
are well known to produce fluid flows, they are attached to the surface,
and external control has only been achieved for synthetic variants
\citep{MASUDA:2016:ID965}. We show that the force produced by the
collective motion of \emph{T. aceti} is substantial and sufficient
to change the evaporation mode of a droplet in which they swim. This
let us anticipate the possibility of designing channels with controlled
flows as well as objects pushed by the nematodes.

\section*{Characterization of the collective state}

We grow and study \emph{T. aceti} in a 1:1 solution of water and apple
cider vinegar. To investigate the collective motion\emph{,} we put
droplets of high density solution of nematodes ($d\geq10\thinspace\mathrm{n/\mu l}$)
on glass slides whose surface was treated with a hydrophobic PDMS
compound. Most experiments were done with 100 μl droplets, though
we verified that similar behavior can be observed in droplets up to
1000 μl and down to 50μl. Additional information on the preparation
and experiments is available in the supplemental material SII.

\begin{figure*}
\includegraphics[width=1\textwidth]{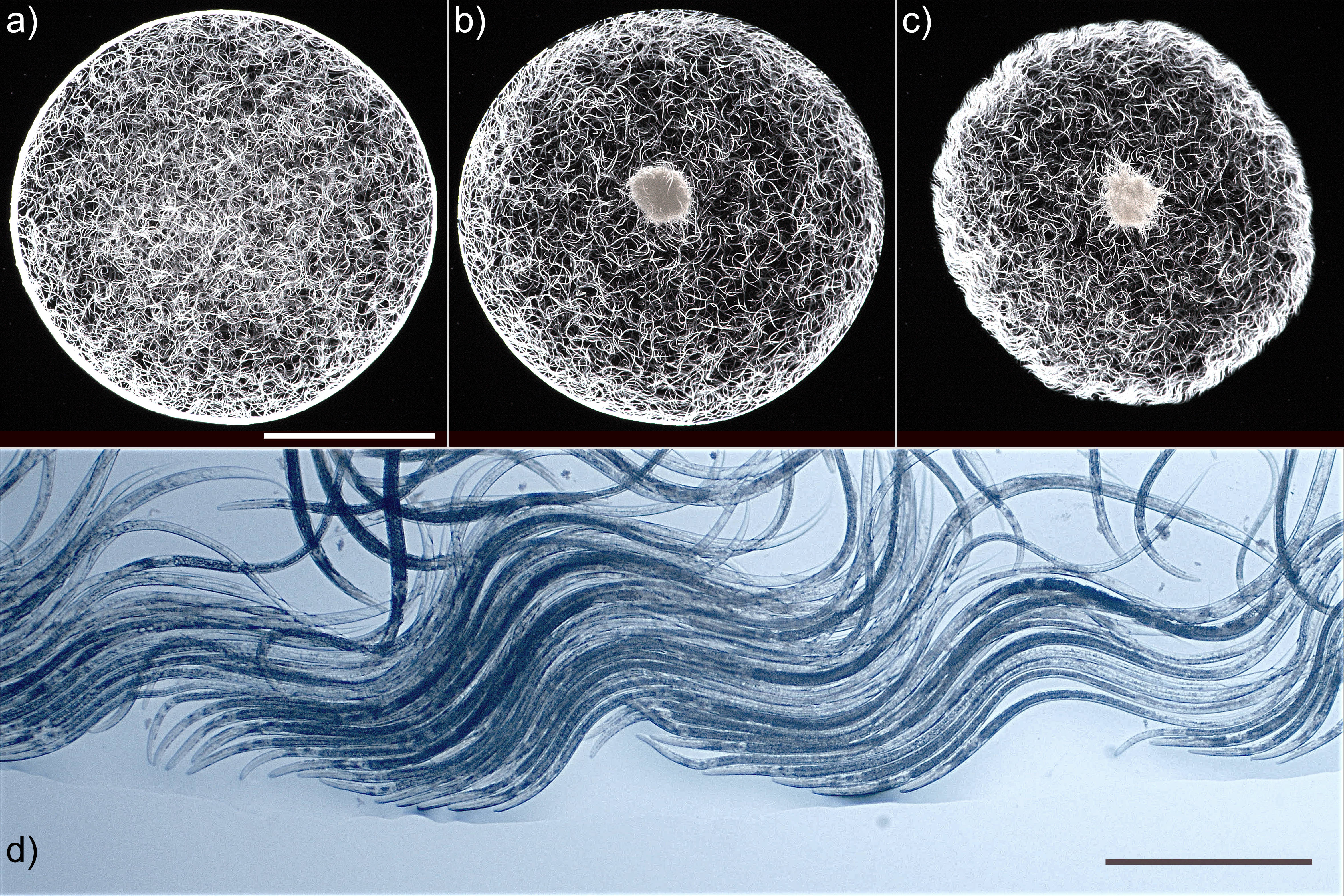}

\caption{\label{fig:Droplets-photo}Top: a-c) Photos of evaporation of a 250
μl droplet at different moments in time. Initial density of nematodes
in the droplet was $d=30.7\pm3.3\mathrm{n/\mu l}$. Scale bar is 5
mm. a) t=1 min, random motion. b) t=20 min, percolation of the metachronal
wave at the border. c) t=60 min, fully developed metachronal wave.
d) A view of the metachronal wave similar to that in c) under a microscope
with 4x magnification. Scale bar is 0.5 mm.}
\end{figure*}

Initially the motion of nematodes in the droplet is random as can
be seen in Figure \ref{fig:Droplets-photo} a) and supplemental movie
SM1 (movies are available at\citep{PESHKOV:2021:ID1057}). After the
deposition on glass, the nematodes start to concentrate on the border
of the drop due to bordertaxis \citep{Yuan_2015}. Individual nematodes
that approach the border continue moving along the border as expected
for active particles. After a variable period of time, depending on
the droplet volume and evaporation conditions, the oscillation of
groups of nematodes on the border becomes locally synchronized. If
the concentration of nematodes in the initial droplet is large enough,
with number of organisms per unit volume $d_{wave}\gtrsim10-20\mathrm{n/\mu l}$,
the locally synchronized swarms will grow in size until finally percolating
into a metachronal wave that spans the whole border of the droplet,
as represented on Figure \ref{fig:Droplets-photo} b) and supplemental
movie SM2. The number of nematodes participating in the wave and the
degree of synchronization increase in time until reaching a maximum
wave strength as illustrated in Figure \ref{fig:Droplets-photo} c)
and supplemental movie SM3. To the best of our knowledge, this is
the first report of such collective motion in this, or any other specie
of nematodes. The existence of this, both swarming and oscillating,
state is the first major finding of this manuscript.

Observing this collective wave under a microscope (Figure \ref{fig:Droplets-photo}
and supplemental movie SM4), we can see that the nematodes orient
their head toward the border and synchronously oscillate at an angle
from 0 to $90\text{°}$. While nematodes of different ages and sizes
are present in the solution, only similarly sized adult ones participate
in the wave. Smaller nematodes are expelled to the center of the drop.
We were able to observe wave frequencies $f_{wave}$ in the range
of 4-8 Hz, consistent with that of individual nematodes. Though the
majority of experiments produced waves with lower frequencies in the
4-5 Hz range. While the mechanism of selection of beating frequency
needs to be elucidated, once the wave is formed, the frequency remains
constant for the lifetime of the wave. In addition to their oscillations,
the nematodes slowly move along the border with a typical velocity
of $v_{border}\approx0.1\mathrm{mm/s}$. This velocity is lower than
that of freely swimming nematodes $v_{swim}\approx0.4\mathrm{mm/s}$
and an order of magnitude smaller than the velocity of the phase of
the metachronal wave $v_{wave}\approx4\mhyphen6\mathrm{mm/s}$. Note
that in this state, the position and the velocity of the nematodes
is controlled by the oscillations of the nearby nematodes. In that
sense, this state is a realization of \emph{swarmalotors.}

In more than 70\% of the experiments that we have performed, the wave
rotates in the counter-clockwise direction. We do not have an explanation
for this symmetry breaking. A similar symmetry breakage was observed
in colonies of rotating magnetotactic bacteria under the influence
of a magnetic field \citep{VINCENTI:2019:ID943}. It was speculated
that this absence of symmetry could be due to the helicity of the
bacteria. Therefore, this skewness of rotational direction of the
wave that we observe could indicate a possible asymmetry in the motion
of individual swimming nematodes, though we were not able to observe
it in practice \citep{QUILLEN:2021:ID1073}.

In addition to the metachronal wave, we often observe the formation
of a compact cluster of nematodes in the vicinity of the center of
the droplet and a strong reduction in the density of nematodes in
the space between the border and the dense cluster at the center (see
Figure \ref{fig:Droplets-photo}b) and c)). When observed under the
microscope, the cluster presents itself as a knot of highly entangled
nematodes similar to the ones recently observed for \emph{C. elegans
}on solid surfaces \citep{DEMIR:2020:ID901} as well as \emph{T. tubifex}
\citep{DEBLAIS:2020:ID906} and \emph{L. variegatus} \citep{OZKAN-AYDIN:2021:ID1087}
in liquid. This cluster is not stable in time and can grow and shrink
during the experiment. The dynamics of these clusters is out of the
scope of this article and will be presented in future works.

When the wave is formed, it is generally stable until the end of its
life. However, instabilities can be observed for large droplet sizes
(>500 μl) and in droplets with low contact angles (< 30 °). A temporal
increase in size of the central cluster can lead to the depletion
of nematodes at the border and the disappearance of the wave. More
information on instabilities is provided in supplemental part IV.

If the contact angle of the drop is low enough, a strong deformation
of the border of the drop will occur, as can be seen in the supplemental
movie SM5. This indicates the existence of strong currents produced
by this collective state. Indeed we were often able to observe a rotational
motion of the free swimming and clustered nematodes, which were not
part of the wave, as well as of tracer particles, in the center of
the droplet, as illustrated by the supplemental movie SM6. In the
mentioned movie, we measured the rotational velocity in the center
of the drop to be around 1.3 rpm. Given that the fixed quantity is
the above-mentioned metachronal wave travel velocity $v_{wave}$,
the rotational velocity will decrease in larger droplets and will
increase in smaller ones.

\section*{Conditions for the collective state}

We are interested in identifying parameters that control the appearance
of the metachronal wave during the evaporation of the droplet. One
hypothesis would be that the metachronal wave appears when the density
of nematodes reaches a threshold value. However, as mentioned before,
we see the appearance of metachronal wave for all the droplets as
soon as the initial concentration of nematodes is sufficiently large
$\apprge10-20\mathrm{n/\mu l}$. Neither the timenor the size of the
droplet is a determining factor as we show below. We have discovered
that the control parameter for the formation of the metachronal wave
is the drop contact angle.

\begin{figure}
\includegraphics[width=1\columnwidth]{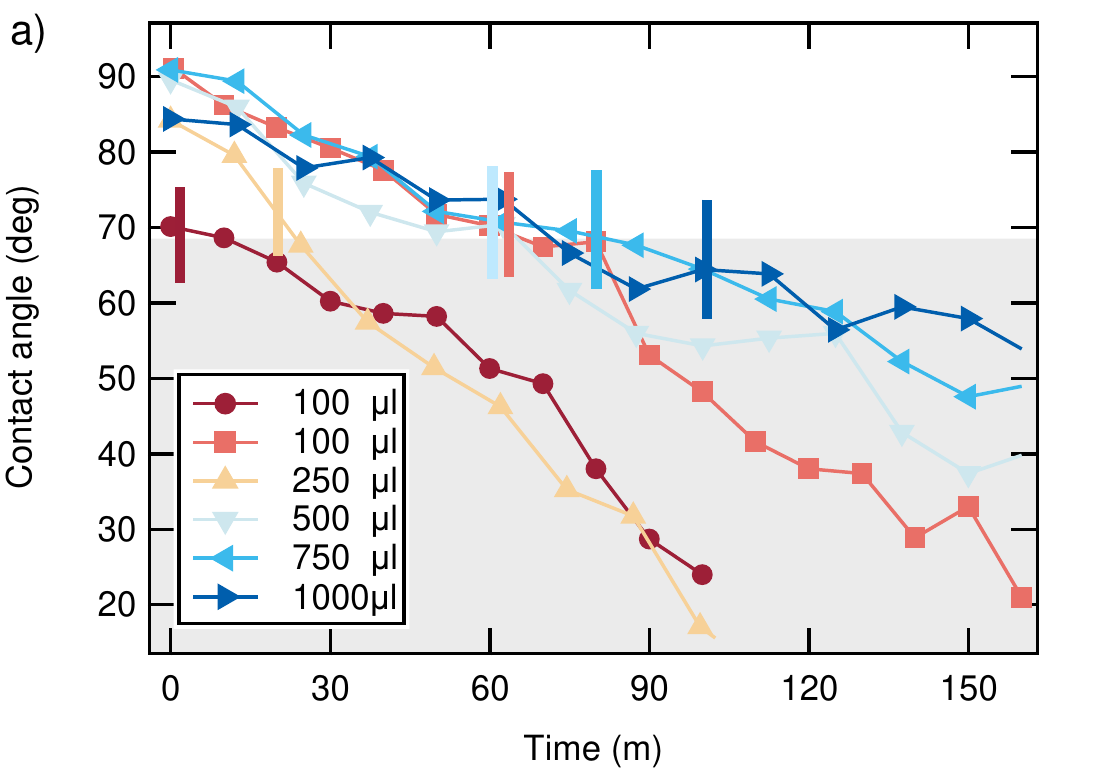}

\includegraphics[width=1\columnwidth]{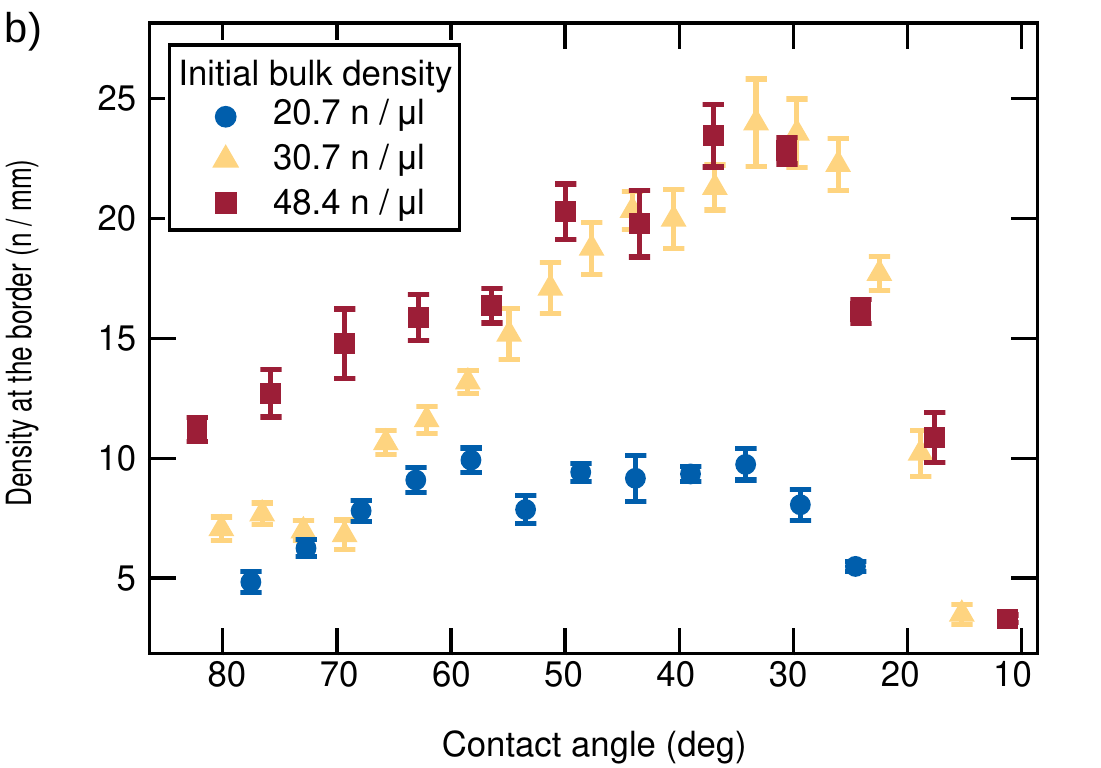}

\includegraphics[width=1\columnwidth]{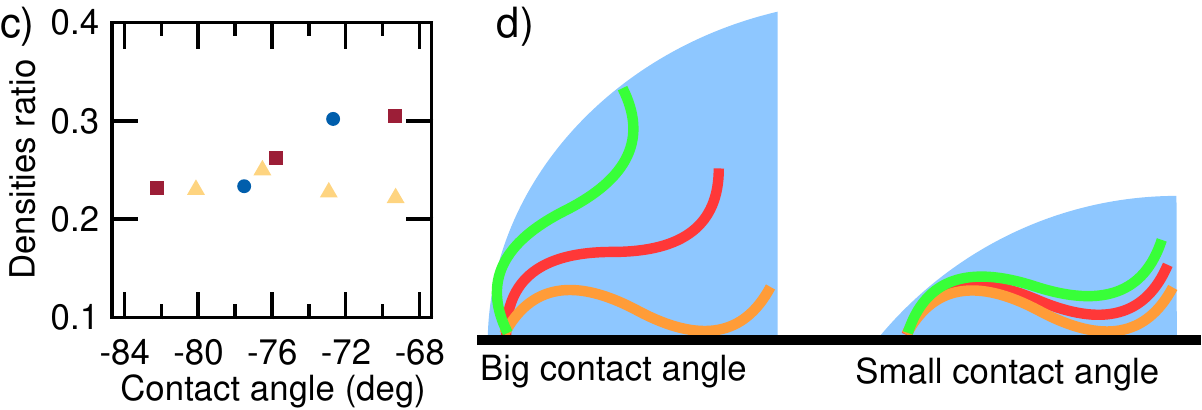}

\caption{a) Contact angle of evaporating droplets over time. Different lines
represent different experiments with different initial volumes and
concentrations, as well as different evaporation conditions. The 250
μl droplet is the same as Figure \ref{fig:Droplets-photo}a)-c). Colored
vertical lines represent the approximate time when the full drop spanning
metachronal wave was formed. The shadowed gray area corresponds to
the angles where the metachronal wave exists. The estimated standard
deviation on angle determination is $4\text{°}$. b) Density of nematodes
at the border as a function of the contact angle for three different
initial density of nematodes in the solution. No metachronal wave
was observed for the lowest density. Error bars are standard mean
error. c) Ratio of border to bulk density for contact angles superior
to $\theta_{c}$. d) A possible explanation of the dependence of the
collective state on the droplet contact angle. At high contact angle
(left), the nematodes are unlikely to touch each other. At low contact
angle (right), the nematodes will touch each other and therefore synchronize
their motion.\label{fig:Contact-angle-single-density}}
\end{figure}

Figure \ref{fig:Contact-angle-single-density} a) shows the contact
angle of several selected droplets as a function of time. We selected
several representative droplets of different densities, volumes, initial
contact angles and evaporation rates. We indicate with a vertical
bar, on each curve, the time at which the metachronal wave spanning
the whole drop perimeter appears. We can see that time is not a determinant
parameter for the formation of the collective state, with times ranging
from several seconds to several hours. However, the metachronal wave
appears for all the droplets at approximately the same contact angle.
We have measured the critical angle for the percolation of the metachronal
wave to be $\theta_{c}=68.5\pm1\lyxmathsym{\textdegree}$.

This phase transition to the collective state at a particular angle
is confirmed by analyzing the number of oscillating nematodes at the
border of the droplet $d_{b}$ (Figure \ref{fig:Contact-angle-single-density}
b). For droplet contact angles superior to $\theta_{c}$, the density
of nematodes at the border is mostly independent of the angle and
is directly proportional to the density of nematodes in the initial
solution (Figure \ref{fig:Contact-angle-single-density} c). However,
past the critical angle the density of nematodes start to increase.
In the case of an initial density insufficient to form a full wave,
we see only a small increase in density up to an angle of around $60\text{°}$,
and staying constant afterwards. In contrast, for droplets in which
a collective wave appears, the density at the border grows until reaching
a maximum at an angle close to $35\text{°}$. The maximum density
at the border seems to be independent of the initial concentration
of the solution. This can be easily explained, as the maximum instantaneous
density that we were able to observe was $29.9\thinspace\mathrm{n/mm}$.
Given that the typical width of a single nematode participating in
the wave is $29(5)\thinspace\mathrm{\mu m}$, this allows us to estimate
the densest ``close-packing'' at $34.5\left(6.0\right)\mathrm{\thinspace n/mm}$,
for nematodes perpendicular to the surface. However the nematodes
bodies are oscillating at a tilt of $\sim20\text{°}$\citep{QUILLEN:2021:ID1073},
which limits the maximum density to $\sim32.4\thinspace\mathrm{n/mm}$,
slightly higher than the value we measured. For droplet contact angles
inferior to $30\text{°}$, we observe a sharp decrease in the number
of the nematodes on the border and a disappearance of the collective
wave. This can be explained by the space constraint which does not
allow the nematodes to effectively fit into the fluid volume near
the border . The dependence of the location and strength of the metachronal
wave on the contact angle of the droplet is the second major finding
of this manuscript.

A natural question is why the formation of the wave would depend on
the contact angle. Our hypothesis is represented in Figure \ref{fig:Contact-angle-single-density}
d). In \citep{QUILLEN:2021:ID1073} we suggested that the synchronization
of nematodes is facillated by steric interactions between them. If
the contact angle of the droplet is high, the probability that two
nematodes oscillating at the border and located nearby one another
will touch is relatively low, as they will in most cases oscillate
at different angles to the surface, making synchronization less likely.
In the opposite case, if the contact angle of the droplet is low and
the drop is very shallow, two nematodes oscillating nearby will almost
certainly touch each other and strongly interact and synchronize their
motion. Note that the synchronization of the motion of nematodes,
will ``free-up'' the space at the border, as they can be closer
to one another. This explains the increase in concentration near the
border as the angle of the droplet decrease, and the wave becomes
more synchronized. It is also important to remark that, the concentration
of nematodes at the border is dependent on the contact angle, but
not the ``history'' of achieving this angle as illustrated by the
droplet that started at an angle of $70\text{°}$ on Figure \ref{fig:Contact-angle-single-density}
a). Thus the bordertaxis is not due to any outward flows due to droplet
evaporation like in the ``coffee-ring'' effect, as opposed to for
example the motion of microalgae Chlamydomonas reinhardtii inside
droplets \citep{BITTERMANN:2021:ID1085}.

\section*{Droplet evaporation}

\begin{figure}[h]
\includegraphics[width=1\columnwidth]{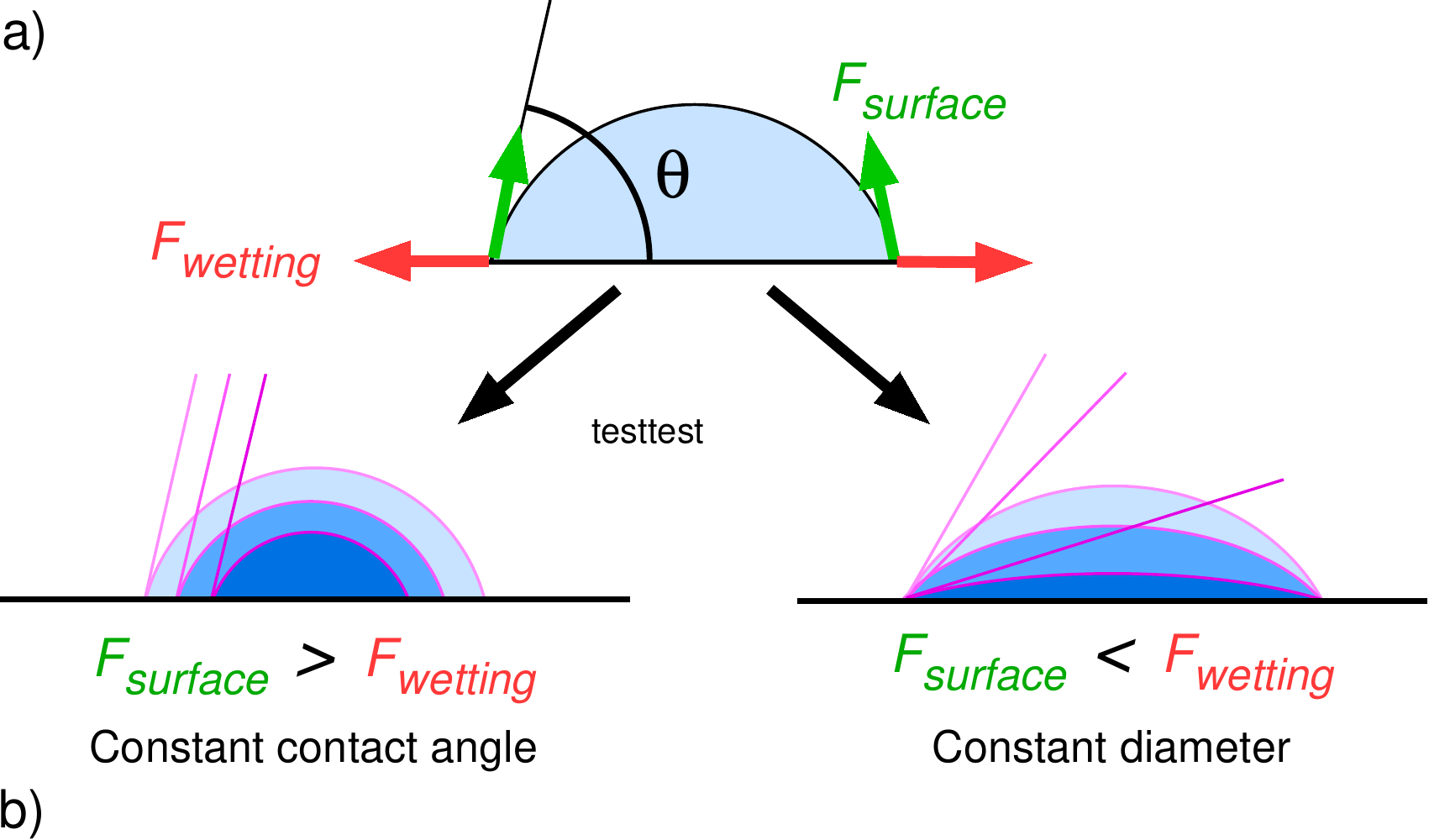}

\includegraphics[width=1\columnwidth]{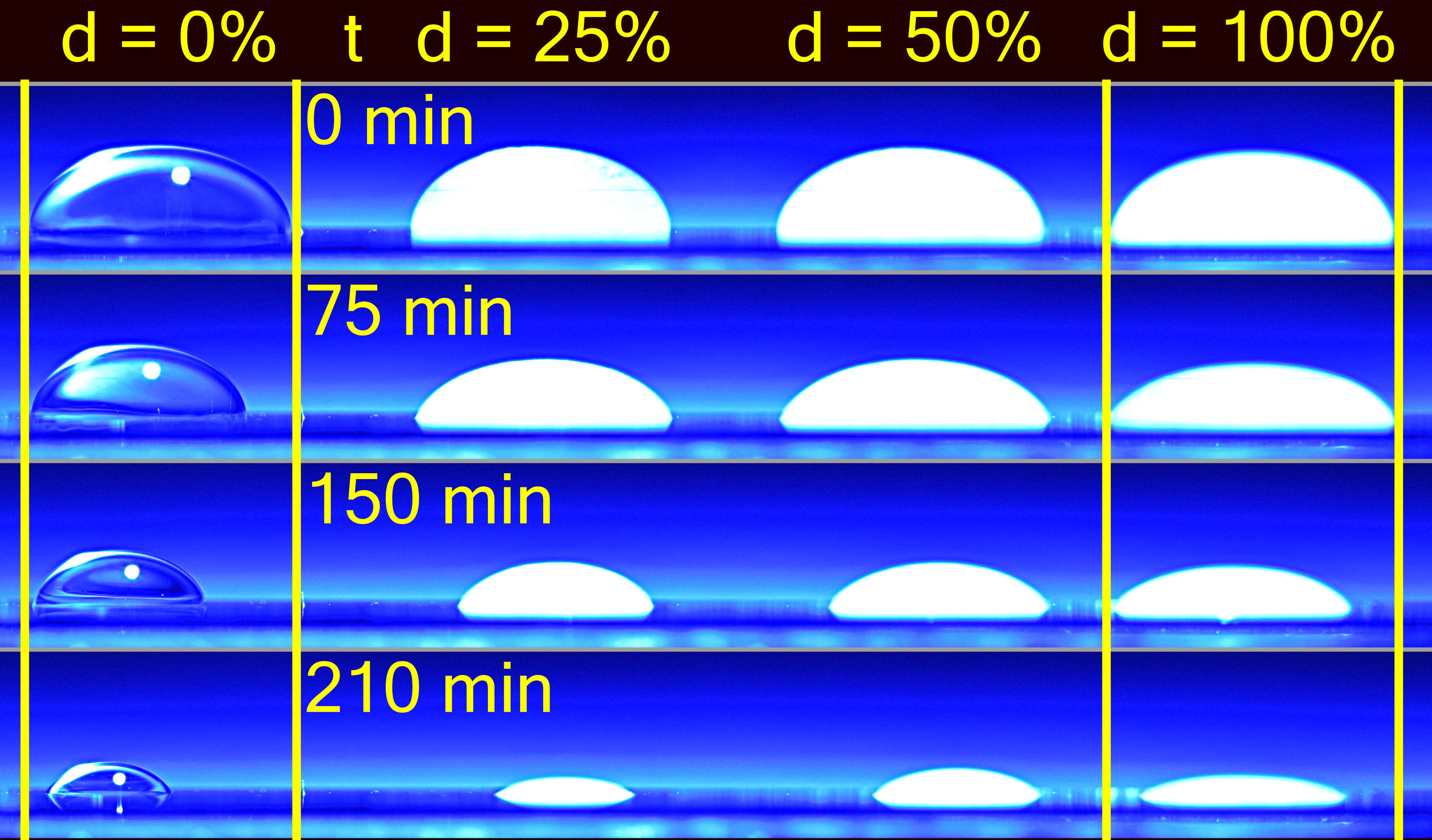}

\caption{a) Diagram of two different possible modes of drop evaporation depending
on the force balance between the fluid and the surface. b) Side view
of four 100 μl droplets with different nematode concentrations during
evaporation. The concentrations of nematodes are from left to right
0\%, 25\%, 50\%, 100\%, with the rightmost drop taken as a reference
density. Vertical yellow lines show initial borders of the 0\% and
100\% concentration droplets.\label{fig:Miltiple-densities-photos}}
\end{figure}

The physics of drop evaporation is relatively complex and depends
on the properties of the liquid, the surrounding gas and the surface
on which the drop resides \citep{HU:2002:ID950}. The process of drop
evaporation is generally divided into two or three main stages \citep{BOURGES-MONNIER:1995:ID936}.
In the following we will focus on the first stage of evaporation,
neglecting the later stages of very sharp reduction in the droplets
contact angle or diameter. There exist two different modes \citep{PICKNETT:1977:ID946}
of this phase of drop evaporation as shown in Figure \ref{fig:Miltiple-densities-photos}
a). The first, and most common case, notably for water on glass, is
a constant contact surface area, when the drop evaporates through
the decrease of the contact angle. For simplicity we will refer to
this mode as the one at a constant diameter, as diameter is the parameter
that we measure. This mode of evaporation will appear if the wetting
contact forces to the surface are greater than the surface tension
forces. For most common fluids, such as water, the initial contact
angle of the drop will be less than 90° \citep{BIRDI:1993:ID948}.
However, on some hydrophobic surfaces, the wetting force will be less
than the surface tension force. In that second case, the drop will
reduce its surface area while maintaining a constant contact angle
during evaporation, and the drop diameter will decrease. Examples
of such surfaces for water are polytetrafluoroethylene (PTFE, commonly
known as Teflon) and the PDMS that coat our glass slides \citep{BIRDI:1993:ID948}.
For most fluids, including water, the initial contact angle on such
surface will be typically equal or greater than 90° \citep{BIRDI:1993:ID948}.

Since our slides were covered with hydrophobic PDMS, we will expect
the contact angle to remain constant during evaporation. However we
have seen in the previous section that this was actually not the case
for our droplets with the nematodes. To confirm these, we perform
an experiment with four droplets of different concentrations, prepared
from the same initial dense solution. Because vinegar and other suspended
particles in the grow solution will inevitably affect the droplet
evaporation, for this experiment we transfer the nematodes into distilled
water as described in the methods section. While such a transfer present
a risk of nematodes swallowing and busting due to osmosis\citep{KISIEL:1975:ID951},
we have not observed a change of their behavior in the short time
span of up to four hours of these experiments.

\begin{figure}[h]
\includegraphics[width=1\columnwidth]{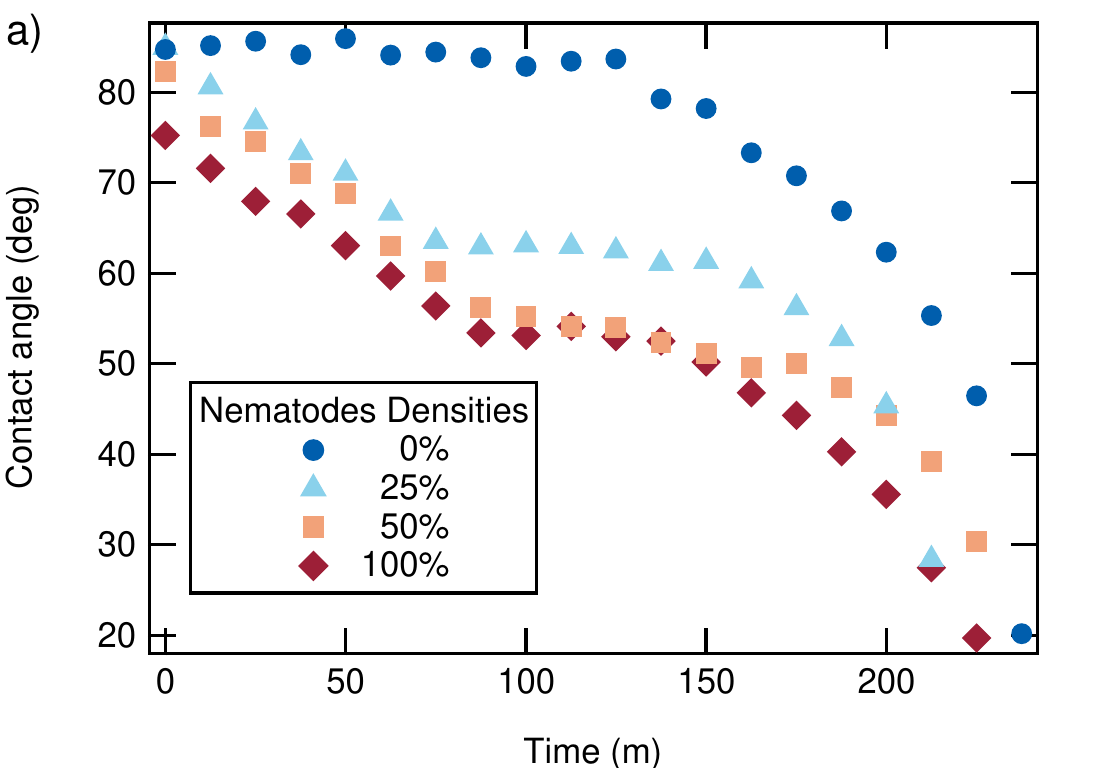}

\includegraphics[width=1\columnwidth]{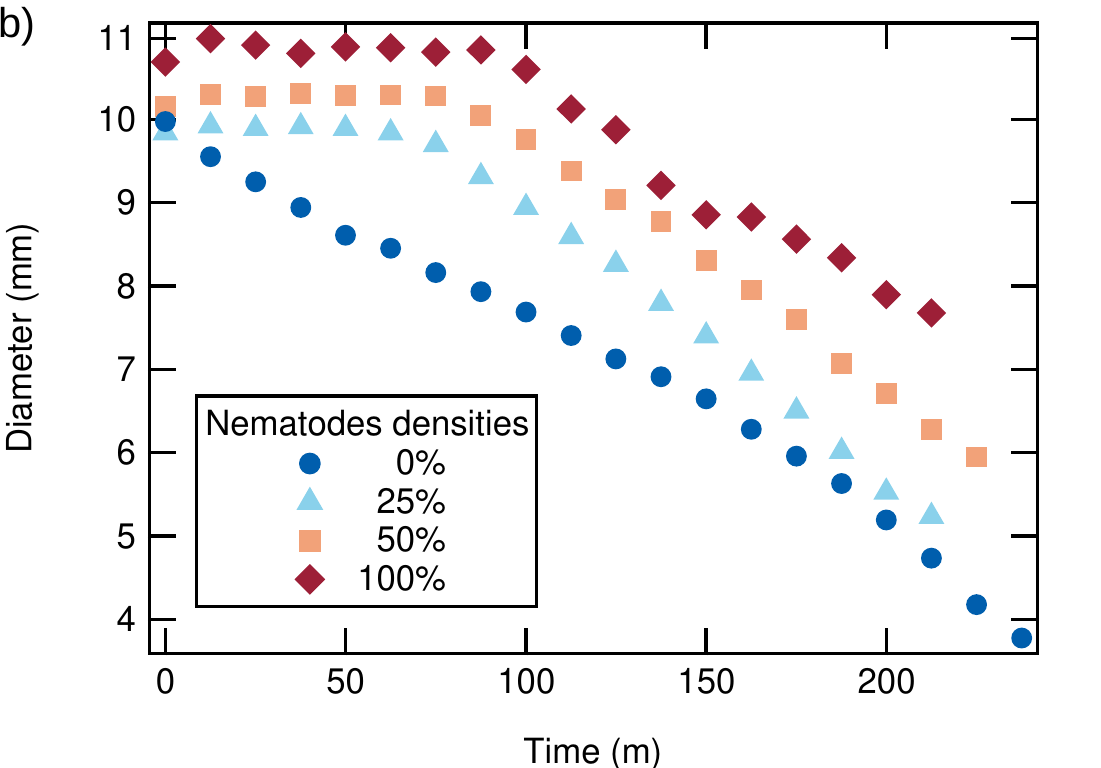}

\caption{\label{fig:multiple-densities-extracted-values}Contact angles (a)
and diameters (b) of the droplets from Figure \ref{fig:Miltiple-densities-photos}
as a function of time. The estimated standard deviation is $1\text{°}$
for the angle and 0.1\ mm for the diameter.}
\end{figure}

Figure \ref{fig:Miltiple-densities-photos} b) shows the side view
of four droplets at different moments of time for four different concentrations
of nematodes: 0\%, 25\% $\left(d=22.5\ \mathrm{n/\mu l}\right)$,
50\% $\left(d=45.1\ \mathrm{n/\mu l}\right)$ and 100\% $\left(d=90.2\pm5.3\ \mathrm{n/\mu l}\right)$.
As expected for water, the 0\% density droplet evaporates by reducing
the diameter. On the contrary, the droplets with the nematodes initially
evaporates with a constant diameter and decreasing angle. Figure \ref{fig:multiple-densities-extracted-values}
shows the extracted contact angles and diameters of these droplets.
The d=0\% droplet evaporates with a constant contact angle and a continuously
decreasing diameter until reaching the second phase of evaporation
when the contact angle starts to rapidly decrease. For the droplets
with the nematodes, we observe a reduction in contact angle until
it reaches a transition angle $\theta_{t}$, which is different for
each droplet. After reaching $\theta_{t}$, the droplets transition
from the constant diameter mode of evaporation to the constant contact
angle mode. Finally, at even later times, the second stage of evaporation
with a sharp decrease in contact angle is observed. Surprisingly,
the diameter of the droplets with the nematodes slightly increase
at the beginning, an observation made by us in many experiments. This
maybe explained by the facilitation of the contact point hysteresis
by the random motion of the nematodes near the border. That the collective
motion of nematodes can affect the physics of droplet evaporation
is the third major finding of this work.

\section*{Force produced by the nematodes}

\begin{figure}
\includegraphics[width=1\columnwidth]{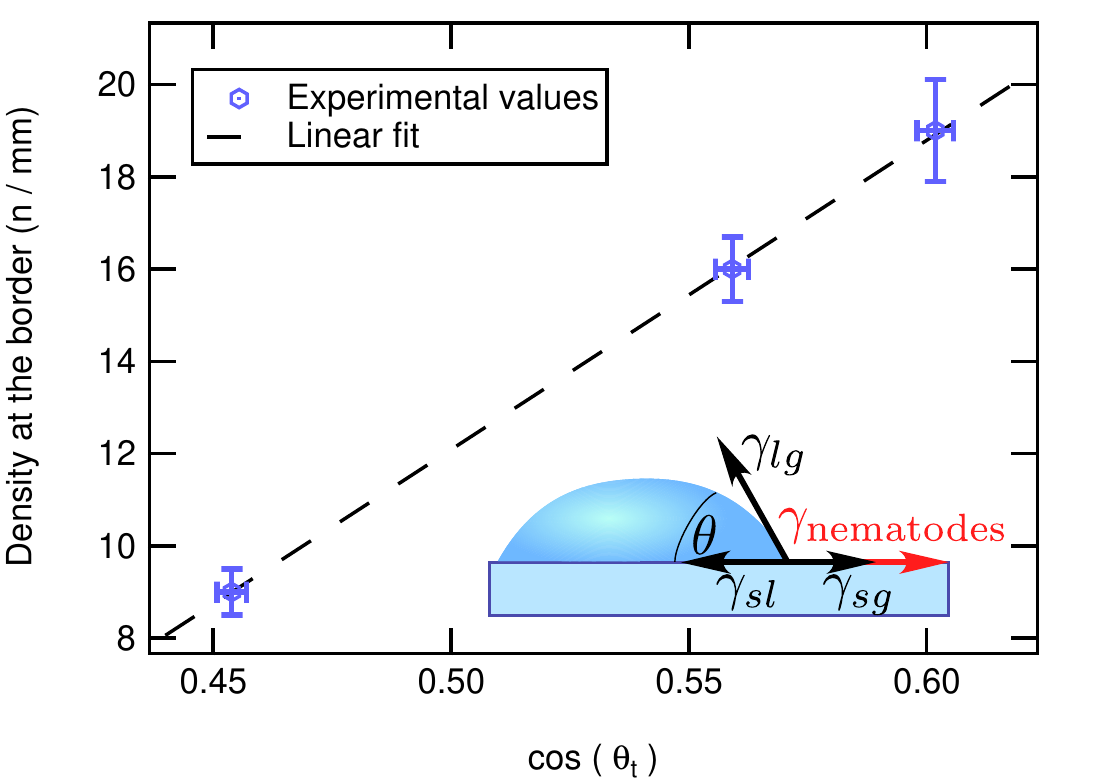}

\caption{\label{fig:Force-fit}Expected density at the border versus the cosine
of the transition angle $\theta_{t}$ for the three droplets from
Figures \ref{fig:Miltiple-densities-photos}, \ref{fig:multiple-densities-extracted-values}.
Inset: Diagram of forces acting on the contact point in a droplet
residing on a flat surface.}

\end{figure}

To explain the existence of the transition angle $\theta_{t}$ , let
us go back to the physics of droplets evaporation. The balance between
the different forces acting on a droplet in contact with a flat solid
surface is given by the Young's relation
\begin{equation}
\gamma_{sg}=\gamma_{sl}+\gamma_{lg}\cos\theta.
\end{equation}
Here $\gamma_{sg},\gamma_{sl},\gamma_{lg}$ are the surface tensions
of solid/gas, solid/liquid and liquid/gas interfaces and $\theta$
is the contact angle as explained in the inset of Figure \ref{fig:Force-fit}.
In simple words, $\gamma_{lg}$ measures how much the liquid and ambient
air molecules repel one another; $\gamma_{sl}$ how much the liquid
molecules repel the solid ones; and $\gamma_{sg}$ how much the solid
molecules repel the surrounding gas. Therefore, $\gamma_{lg}$ and
$\gamma_{sl}$ tend to decrease the diameter of the droplet (assuming
$\theta<90\text{°}$), while $\gamma_{sg}$ tend to increase it.
In addition to these forces, the nematodes could exert and outward
horizontal force $\gamma_{n}$ on the boundary. We modify Young's
relation to include this force per unit length giving 
\begin{equation}
\gamma_{sg}+\gamma_{{\rm n}}=\gamma_{sl}+\gamma_{lg}\cos\theta.
\end{equation}

Note that as the angle $\theta$ will decrease, the horizontal force
on the contact point coming from the surface tension contribution
$\gamma_{lg}$ will become stronger. Thus at some critical angle,
the force from the nematodes $\gamma_{n}$ would not be able any longer
to oppose the surface tension force, and the mode of evaporation will
change from that at constant diameter to the one at a constant contact
angle. To verify this hypothesis let us assume that the force exerted
by the nematodes is proportional to the density of the nematodes at
the border $\gamma_{{\rm n}}=F_{n}d_{b}$, where $F_{n}$ is the force
exerted by an individual nematode and $d_{b}$ the density of nematodes
at the border that we have previously measured. Our Young's relation
at the transition angle then becomes

\begin{equation}
F_{n}d_{b}=\gamma_{sl}-\gamma_{sg}+\gamma_{lg}\cos\theta_{t}.
\end{equation}

The three values of the transition angle in Figure \ref{fig:multiple-densities-extracted-values}
a) are $\theta_{t}\left(25\%\right)=63\text{°}\pm0.15$, $\theta_{t}\left(50\%\right)=56\text{°}\pm0.25$
and $\theta_{t}\left(100\%\right)=53\text{°}\pm0.27$. The densities
at the border for the two smallest density droplets can be directly
taken from Figure \ref{fig:Contact-angle-single-density} c) with
$d_{b}\left(25\%\right)=9\pm0.5\ \mathrm{n/mm}$ and $d_{b}\left(50\%\right)=16\pm0.7\ \mathrm{n/mm}$.
While we have not directly measured the border density of droplets
at such a high concentration as of our 100\% droplet, we can reasonably
assume that because of the final possible packing, this density will
be approximately the same that the one that we measured for our highest
concentration droplet on Figure \ref{fig:Contact-angle-single-density}
c. By linearly interpolating between two points, we therefore obtain
$d_{b}\left(100\%\right)=19\pm1.1\ \mathrm{n/mm}$. Plotting $d_{b}$
versus $\cos\theta_{t}$ we obtain a perfect straight line as predicted
by our theory (Figure \ref{fig:Force-fit}).

The value of $\gamma_{lg}=72.59\left(0.36\right)\mathrm{\mu N/mm}$
\citep{VARGAFTIK:1983:ID1081} for water in contact with air at $21\text{°}C$
temperature is well known and allows us to extract from the fit the
force exerted by an individual nematode on the border $F_{n}=1.08(0.11)\ \mathrm{\mu N}$.
This in turn allow us to compute $\gamma_{sl}-\gamma_{sg}=19.9\left(5.0\right)\ \mathrm{\mu N/mm}$.
While we do not know exactly the composition of the PDMS coating that
we used, the numerically and experimentally estimated values at room
temperature of different varieties of PDMS in contact with vapor are
$\gamma_{sg}\approx20-23\ \mathrm{\mu N/mm}$ and PDMS in contact
with water $\gamma_{sl}\approx40-42\ \mathrm{\mu N/mm}$ \citep{ISMAIL:2009:ID1075}.
This gives $\gamma_{sl}-\gamma_{sg}\approx17-22\ \mathrm{\mu N/mm}$,
in excellent agreement with our measurements, which further validate
our theory and hypothesis.

Contrary to the nematodes in water, no transition to a constant contact
angle evaporation can be observed on Figure \ref{fig:Contact-angle-single-density}
a) for the nematodes in the growth solution. Indeed, the surface tension
of a mixture of acetic acid and water at 5\% concentration and $21\text{°}C$
temperature is $\gamma_{vinegar}=51.8\left(0.5\right)\ \mathrm{\mu N/mm}$
\citep{ALVAREZ:1997:ID1082}, smaller than that of water. This alone
would have changed the transition angles above to $\theta_{t}\left(25\%\right)=55.1\text{°}$,
$\theta_{t}\left(50\%\right)=44.1\text{°}$ and $\theta_{t}\left(100\%\right)=38.7\text{°}$.
Additionally, the suspended particles in the growth medium should
further oppose the reduction of the droplet diameter by contact line
pinning \citep{PARISSE:1996:ID1084}. Both of this reasons explain,
why a transition to a different mode of evaporation is generally not
observed for droplets of the initial growth solution.

Is the measured force of approximately 1 $\mathrm{\mu N}$ per nematode
reasonable? Forces where directly measured on the similarly sized
\emph{C. elegans }by observing the deviation of micropillars when
pushed by the nematodes bodies and gave values in the range of 5 to
30 $\mathrm{\mu N}$ \citep{GHANBARI:2008:ID1055}. An almost order
of magnitude difference can be explained by two considerations. First,
the nematode is not oriented perfectly perpendicular to the surface,
but at an angle of $\approx20\text{°\ }$, using a part of it's force
to move along the wall. However, such a small angle will provide only
a very small decrease in the force ($\cos\left(20\right)\approx0.94$).
Second, if one looks at Figure \ref{fig:Droplets-photo} d), we can
see that only a small proportion of nematodes touch the border at
a given time, which should further reduce the average force exerted
over time. Finally, we expect that the force associated with forward
motion that we measured, is only a small fraction of that exerted
by the entire nematode on the fluid when swimming. Indeed, we have
seen in the first part of the manuscript, that the motion of the nematodes
is capable of deforming the droplet surface. Such deformation is directly
proportional to the force exerted by the undulating nematode body
on the droplet surface. The force needed to produce this curvature
of the surface can be easily computed as described in the supplemental
material SV, and give an estimated value of $F_{undulation}=11\ \mathrm{\mu N}$,
in good agreement with the forces produced by \emph{C. elegans}.

Note, that such a good accuracy of the hypothesis that the force exerted
by the nematodes is proportional to their density is actually surprising.
We would expect that due to hydrodynamic interactions between the
closely spaced nematodes, the force that they produce will depend
on the spacing between them. Indeed, the measurement of produced flow,
both in experiments on artificial cilia \citep{XIAOGUANG:2020:ID1077}
and simulations of straight paddles moving with metachronal wave motion
\citep{GRANZIER-NAKAJIMA:2020:ID1076}, shows that it depend on the
spacing between the beating organisms. The border density of $d_{b}\approx10-20\ \mathrm{n/mm}$
lead to a spacing between the nematodes of $s_{n}\approx0.05-0.1\ \mathrm{mm}$.
Given the typical length of the nematodes is $L_{n}\approx1\ \mathrm{mm}$,
this provide a relative spacing of $s_{n}/L_{n}\approx0.05-1$, at
which we would expect a noticeable decrease in efficiency with the
reduction of spacing according to the simulations of straight paddlers
\citep{GRANZIER-NAKAJIMA:2020:ID1076} (assuming $Re=0.4$ as explained
in the next section). However, the applicability of a simplistic model
of straight paddles to the much more complex shape of nematodes is
questionable.

\section*{Discussion and conclusions}

We have studied the collective motion of the nematode \emph{T. aceti}
inside droplets deposited on a flat surface. We have shown that if
the concentration of eels is high enough, a metachronal wave will
form on the edge when the contact angle of the droplet is below a
critical value of $\theta_{c}=68.5\text{°}$. This collective state
is an experimental illustration of \emph{swarmalators}, particles
that at the same time align their motion and their oscillations. We
propose that the dependence of the collective wave on the contact
angle of the drop is due to the increased probability of interactions
between the nematodes bodies at low contact angle. The number of nematodes
participating in the metachronal wave increase as the contact angle
of the drop decreases, which can be explained by the fact that synchronization
will increase the free space available at the border. We also show
that the collective motion of the nematodes can change the physics
of drop evaporation. Initial evaporation of droplets containing synchronously
moving nematodes occurs mostly at constant diameter rather than at
constant contact angle. This allows us to estimate the magnitude of
the force exerted by self propulsion of the nematode on the border
at $F_{n}=1.08\ \mathrm{\mu N}$. This is the force of interest for
possible use of nematodes for displacing objects. Conversely, the
force produced by the undulation of the nematodes bodies can be estimated
from the deformation of the drop surface and is $F_{undulation}=11\ \mathrm{\mu N}$,
in good agreement with direct measurements of the force produced by
the specie \emph{C. elegans}.

Another novelty of our system stems from the relatively large length
of our nematodes, $L_{n}\sim1\mathrm{mm}$, which combined with a
typical swim velocity $v_{swim}\approx0.4\mathrm{mm/s}$ \citep{QUILLEN:2021:ID1073}
leads to a characteristic Reynolds number of $Re=0.4$. This places
our system in the intermediate Reynolds number regime \citep{KLOTSA:2019:ID881}
in contrast to the low Reynolds number regime in which most microorganisms,
which exhibit collective or synchronous motion, reside \citep{PURCELL:1977:ID718,Gueron_1998,Vilfan_2006,Uchida_2011}.
This means that the inertia in the motion of the nematodes cannot
be neglected. The full Navier-Stokes equations should be used to describe
such systems, which become non-linear and time dependent, and thus
can lead to many interesting new states \citep{CHATTERJEE:2019:ID911}.

To better understand what drives the formation of the collective state
it could be interesting to be able to genetically choose the properties
of the nematodes such as sensitivity to light or touch, in the same
way as it is done for the much more studied \emph{C. elegans}. While
it is in theory possible to apply the same genetic toolkit to \emph{T.
aceti}, this will certainly be a major undertaking. For this reason,
we tried to reproduce the collective states of \emph{T. aceti} in
suspensions of \emph{C. elegans}. However, as described in supplemental
material part VI, we were unable to observe any synchronization of
oscillations of \emph{C. elegans} for the various densities and droplet
shapes that we tried. This absence of synchronization may be explained
by the shorter bodies of \emph{C. elegans.}

We believe that \emph{T. aceti} is an extremely promising organism
both for exploring the behavior of \emph{swarmalators }and the states
of active matter at intermediate Reynolds numbers. Much of the physics
of this nematode remains to be explored; the nature of the phase transition
to collective motion, the formation of clusters, its behavior in liquids
of different viscosity or inside confined spaces. Given that the theoretical
exploration of the motion of the nematode will require the study of
the full Navier-Stokes equations, this could lead to new developments
in numerical and theoretical approaches. As we have shown in this
article, the collective motion of the nematode produces strong fluid
flows. As we have an external control parameter for the collective
motion, in the form of the contact angle, we may in the future produce
on-demand flows using specially designed channels. \emph{T. aceti}
combine ease of culture and experimentation with extremely interesting
physics. We hope that this article will start a new thriving direction
of research in the field of active matter.

\section*{Data accessibility}

The original data for this manuscript is available at \citep{PESHKOV:2021:ID1057}.

\section*{Authors' contributions}

A.P. designed the experiments and drafted the manuscript. A.P. and
S.M. performed the experiments and analyzed the data. A.P. ad A.Q.
conceived the study, performed analytical computations and edited
the manuscript. All authors discussed the results and gave final approval
for publication.

\section*{Funding}

This work was supported by NASA grants 80NSSC17K0771 and 80NSSC21K0143;
and National Science Foundation grants No. PHY-1757062 and No. DMR-1809318.

\section*{Competing interests}

The authors have no competing interests to declare.
\begin{acknowledgments}
We thank Myron W. Culver for giving us access to a wet lab. We thank
William Houlihan for lending us an inverted microscope. We thank Nick
Reilly for help procuring a centrifuge. We thank Doug Portman for
helpful discussions on \emph{C. elegans}. We thank Keith Nehrke, Sanjib
K. Guha, Yunki Im and other members of Nehrke's lab for helping us
explore \emph{C. elegans}, and giving us materials and starting cultures
to study \emph{C. elegans}. We thank Steve Teitel and Hugues Chaté
for their careful reading and critique of the earlier versions of
this manuscript. We thank Sanjib K. Guha, Keith Nehrke and Randal
C. Nelson for helpful suggestions and discussions. 
\end{acknowledgments}

\bibliographystyle{apsrev4-2}
\bibliography{anton_references,alice_references}

\end{document}


\title{Synchronized oscillations in swarms of nematode Turbatrix aceti\emph{}\\
Supplemental material}
\author{Anton Peshkov}
\affiliation{Department of Physics and Astronomy, University of Rochester, Rochester,
NY 14627, USA}
\author{Sonia McGaffigan}
\affiliation{Department of Physics and Astronomy, University of Rochester, Rochester,
NY 14627, USA}
\author{Alice C. Quillen}
\affiliation{Department of Physics and Astronomy, University of Rochester, Rochester,
NY 14627, USA}
\maketitle

\section{Notations}

Following the customary convention, we use throughout the article
and the supplemental material the $mean\pm se$ notation to indicate
the standard mean error and the $mean(sd)$ notation to indicate the
standard deviation.

\section{Experimental methods}

We obtained the starter cultures of \textsl{T. aceti} from two different
aquarium supplies stores, to verify that the observed behavior wasn't
particular to a specific strain. We grow these populations in a 1:1
solution of water and food grade apple cider vinegar at $\sim5\%$
acidity in which we put slices of apples as a food source. The population
of nematodes in the culture reaches a peak density after one to two
months, and stays relatively constant afterward. We have observed
that \emph{T. aceti} can successfully survive for many days with limited
access to food and oxygen. The reproduction cycle of \emph{T. aceti}
takes many days and they can have a lifespan of up to two months \citep{KISIEL:1975:ID951}.
Neither the motility nor the number of nematodes significantly varies
for the duration of our experiments of a few hours.

We do not perform any generation control, which mean that nematodes
at different stages of development, and therefore size, are present
in the studied solutions.

For each experiment, 7-14 ml of nematode culture at peak density,
is centrifuged for 3-5 minutes at an acceleration of 1700-4700 g.
The concentrated blob of nematodes at the bottom of the tube is extracted
and mechanically separated with a pipette to obtain a homogeneous
high-density solution of nematodes. The studied densities \emph{d}
in the initial droplet ranged from 10 to 80 nematodes per \textgreek{m}l.
We study these high density droplets by depositing them on a glass
slide with volumes in the range of 50-1000 \textgreek{m}l. Below the
smallest 50 \textgreek{m}l volume, the droplet radius is similar to
the length of a single nematode. Therefore the number \emph{N} of
nematodes in each droplet is in the range of 1-100 thousand organisms.
For experiments in distiled water, the blob of nematodes from the
first centrifugation was transfered to a tube with distiled water
and centrifuged again. Therefore we estimate the concentration of
vineager in this solution to be of the order of $0.1-0.2\%$.

For all the experiments we have coated the glass slide with a hydrophobic
solution of Rain-X which contains polydimethylsiloxane (PDMS) as the
main active ingredient. First, this allows the drops to have an initial
contact angle around 90°. This enables study of the whole range of
possible contact angles during evaporation. Secondly, the shape of
the droplet on a hydrophobic surface is very close to circular, facilitating
measurement of the drop diameter and contact angle. In contrast, the
contact line can be irregularly shaped on wettable surfaces.

The experiments were performed in a room with a controlled temperature
of 21(1) °C and humidity 15(5)\%. However, the temperature in the
vicinity of the droplet, and therefore evaporation rate, is mostly
dependent on the used illumination. For the Figure 1 of the main article,
the droplet was illuminated from the bottom for better contrast and
homogeneity of the image. However, for experiments where the determination
of the droplet contact angle was important, the droplet was illuminated
from top/side, to clearly delimit the border. While it is known that
nematodes such as \emph{C. elegans} are sensitive to light \citep{BURR:1985:ID899},
we did not observe any dependence of the results presented in this
article to the intensity or placement of the light source except for
cases of extremely bright and close light sources. In the latter case,
we observed the concentration of nematodes on the side opposite to
the light source. However we believe that the nematodes avoided heat
rather than light.

\section{Data analysis}

\subsection{Drop shape determination}

The relatively large size of our droplets implies that the gravity
force deforms the droplet surface. A spherical approximation for the
drop shape for the contact angle determination is not accurate. Instead
we adopt an ellipse approximation to measure the contact angle for
the smaller ($\leq100\mathrm{\mu l})$ droplets, using the ImageJ
``Contact angle'' plugin, and B-splines fitting \citep{STALDER:2006:ID949}
for droplets of all sizes, using the ImageJ ``DropSnake'' plugin.
In theory, the axisymmetric drop shape analysis (ADSA) method\citep{LI:1992:ID972}
should give the best results for our droplets as it fits an analytical
equation to the data. In practice the built-in artificial limitations
on acceptable parameters in the ImageJ ``LB-ADSA'' plugin\citep{STALDER:2010:ID974}
were preventing us to obtain a correct fitting in a lot of cases.
We were sometime successful in obtaining good results with the DropUI
program\citep{FAVIER:2017:ID973} which is also based on the ADSA
method. However, the impossibility of manually defining the droplet
surface, and the regular failure of the automatic algorithm to detect
it correctly, precluded us from using this program for analysis of
the majority of our droplets.

\begin{figure}[H]
\begin{centering}
\includegraphics[width=0.5\textwidth]{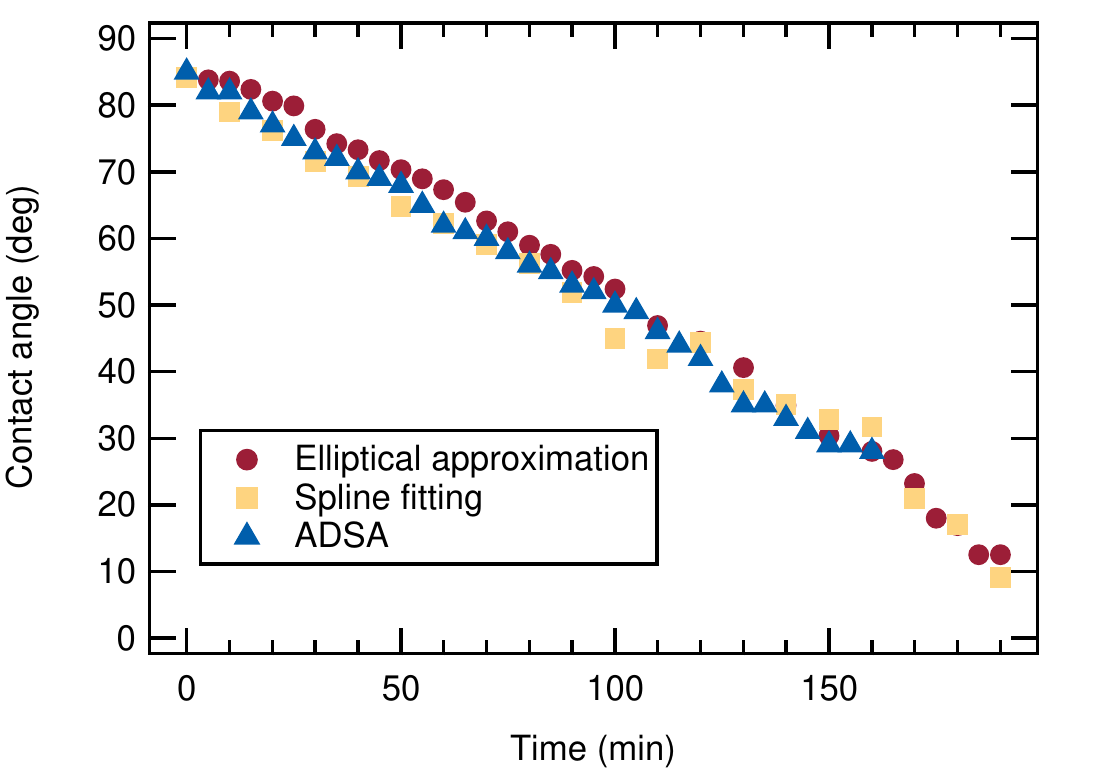}
\par\end{centering}
\caption{Contact angle of a 150 \textgreek{m}l initial volume evaporating droplet
as a function of time as extracted with three different measurement
methods. Red circles, elliptical approximation using the ImageJ ``Contact
angle'' plugin. Yellow squares, B-spline fitting using the ImageJ
``DropSnake'' plugin. Blue triangles, ADSA method using the DropUI
program. \label{fig:Contact-angle-different-methods}}
\end{figure}

Figure \ref{fig:Contact-angle-different-methods} presents a comparison
of the extracted contact angles for the 150\textgreek{m}m droplet
from Figure 2. b) in the main text, using the three approaches described
above. We can see that the elliptical approximation slightly overestimate
the contact angle as compared to the two other approaches. This can
be explained by the relatively large drop size, where the elliptical
approximation is less accurate. Nevertheless, the angle obtained by
the different methods are in reasonable agreement. While spline fitting
give the ``noisiest'' results, it's the method that worked for all
of our droplets and therefore the one we used most frequently. We
note that measuring angles below $\sim20\text{°}$, is very error-prone,
being higly dependent on the allignment of the camera. However, such
small angles are mostly outside of our interest in this article.

From Figure \ref{fig:Contact-angle-different-methods} we estimate
a standard deviation of $4.2\text{°}$ and a standard mean error of
$1.2\text{°}$ for the extracted contact angles in the main article
taking into account the most error-prone elliptical approximation.
This uncertainty does not affect our results or conclusions. Note
that the the angle extraction for the third part of the article was
done using exclusively the ImageJ ``LB-ADSA'' plugin, for which
we estimate a much smaller standard deviation of $1\text{°}$. The
diameter of the droplets were obtained from the droplet surface approximation
by the B-spline fitting of the DropSnake plugin or the ADSA fitting
of LB-ADSA plugin. We estimate a small standard deviation of 0.1\ mm
for the diameter.

\subsection{Nematode Density determination}

To determine the initial density of nematodes, we diluted the dense
nematode suspension 10-50 times in a 50:1 solution of water and glycerol
to prevent droplet diameter shrinkage during evaporation. 10 \textgreek{m}l
droplets of the diluted solution were placed on a glass slide and
evaporated. The number of nematodes in each droplet was then manually
counted under a microscope. This provides the mean and the error estimate
for the initial density.

The number of nematodes per unit length oscillating near the border
in Figure 2. b) was obtained by manually counting the number of oscillating
nematodes near the border in 5-7 frames separated by 5 second intervals
in the vicinity of each contact angle providing the mean and error
estimate.

\section{Instabilities and multiple waves}

\subsection{Wave stability}

In the smaller droplets (<500 \textgreek{m}l), the metachronal wave,
once formed, appears to be stable until its disappearance, when the
drop completely evaporates. In rare cases we observed a temporal dramatic
increase in the size of the central cluster, which leads to the depletion
of the nematodes on the border and a temporal disappearance of the
metachronal wave.

In larger droplets (>500 \textgreek{m}l), we observed that the metachronal
wave can sometimes split into several parts with opposite rotating
directions (see supplemental video SM7) and even fully reverse the
direction of rotation along the whole border. Two possible explanations
can be given to that phenomenon. First, that above some instability
length of the border $L_{i}\approx50\mathrm{mm}$, the wave becomes
unstable due to an instability in the process of synchronization.
Another explanation may be due to external properties of the droplet.
The droplet slightly shrinks during evaporation, and given the non-perfect
nature of the substrate, this shrinkage is not equivalent in all radial
directions. This would lead to the deformation of the border and a
potential nucleation of a ``defect'' in the wave. For larger droplets,
the probability of a substantial deformation of the border is naturally
higher.

\subsection{Multiple waves}

Note that since the amplitude of oscillation for an individual nematode
is approximately  0.1 mm, the nematodes cannot oscillate in the plane
perpendicular to the bottom surface of the drop. Therefore, the oscillations
will be increasingly confined to the bottom surface plane as the contact
angle decreases. This is indeed what we experimentally observe. If
the contact angle of the droplet becomes really small, one may wonder
if the confinement can induce the metachronal wave more distant from
the droplet border. Indeed, for very small contact angles, we sometimes
observed the creation of a secondary wave in the inner radius as can
be seen in the supplemental movie SM8. This wave is always counter-rotating
as compared to the border wave. This is easy to understand, if the
nematodes rotated in the same direction, they will just penetrate
inside the border wave, the only way for such wave to exist is to
rotate in the opposite direction. However, such double waves are relatively
short living, inevitably leading to the perturbation of the outer
wave, and even sometimes to the reversal of the direction of the latter.

\section{Estimation of the force exerted by the nematode on the drop surface}

\begin{figure}[H]
\begin{centering}
\includegraphics[width=0.5\textwidth]{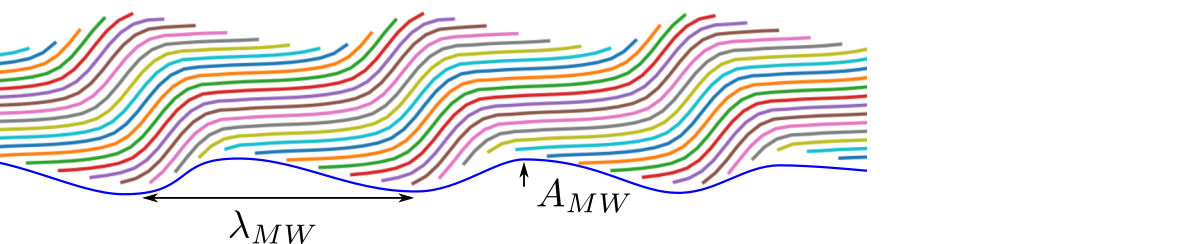}
\par\end{centering}
\caption{\label{fig:wave-diagaram}Diagram of the wave wavelength $\lambda_{W}$
and amplitude $A_{w}$.}

\end{figure}

We have seen that the force produced by the nematodes is sufficient
to deform the surface of the droplet. We can estimate the force on
the surface based on the pressure required for surface deformation
with the wavelength of the metachronal wave. The metachronal wave
has a typical wavelength $\lambda_{w}\sim1$ mm and amplitude $A_{w}\sim0.1$
mm (Figure \ref{fig:wave-diagaram}). We describe the surface with
a height away from the mean $\xi=A_{w}\cos k_{w}x$ where $x$ is
a distance along the surface and wave vector $k_{w}=2\pi/\lambda_{w}$.
The pressure difference can be estimated from the curvature of the
surface 
\begin{align}
\Delta p & \sim\gamma_{lg}\frac{d^{2}\xi}{dx^{2}}=\gamma_{lg}A_{w}k_{w}^{2}\\
 & \sim287\thinspace\mathrm{\mu N/{\rm mm}^{-2}}
\end{align}

To estimate force per nematode we need to estimate the number of them
per unit area near the surface. If we adopt $s_{{\rm nematodes}}\sim25\ {\rm mm}^{-2}$,
the highest mean density which is present for the contact angle at
which we observe surface deformation, the force per nematode is then
given by
\begin{align}
F_{{\rm nematode}} & \sim\frac{\Delta p}{s_{{\rm nematodes}}}\\
 & \sim11\ \mu{\rm N}.\label{eqn:F2}
\end{align}

The two force estimates are similar and consistent with forces measurements
of 5 to 30 $\mu$N exerted by individual \textit{C. elegans} nematodes
that were made using an array of pillars that can bend when pushed
by the nematodes \citep{GHANBARI:2008:ID1055}.

\section{C. elegans}

We sourced the strain \emph{N2} of \emph{C. elegans} grown on agar
plates. We followed the same procedures as described above for \emph{T.
aceti} to try to obtain a collectively oscillating state. We observed
that \emph{C. elegans} exhibit bordertaxis, as already reported in
\citep{YUAN:2015:ID945}, and that individual nematodes oscillate
with their heads oriented to the border in a manner similar to \emph{T.
aceti}. However, we never saw a synchronization of motion between
the nematodes, whatever the nematode concentration and droplet shape
and size we tried. An example is shown in supplemental movie SM9 and
Figure \ref{fig:C.-elegans}. There are two possible explanations.
First, \emph{C. elegans} generally live in soil, and so there may
be no evolutionary advantage in collectively driven fluid flows. Secondly,
the wider and shorter \emph{C. elegans} has less than a full-wavelength
oscillation along the body, and this differs from \emph{T. aceti}
which is longer and more slender. It is known \citep{STIERNAGLE:2006:ID953}
that \emph{C. elegans} grown in a liquid medium, as opposed to those
grown on agar, are longer and thinner, opening up possibilities for
synchronization. However, given the widespread use of \emph{C. elegans
}in medical research, it is likely that if a metachronal wave state
existed for them, it would have already been noticed and reported.

\begin{figure}[H]
\begin{centering}
\includegraphics[width=0.5\textwidth]{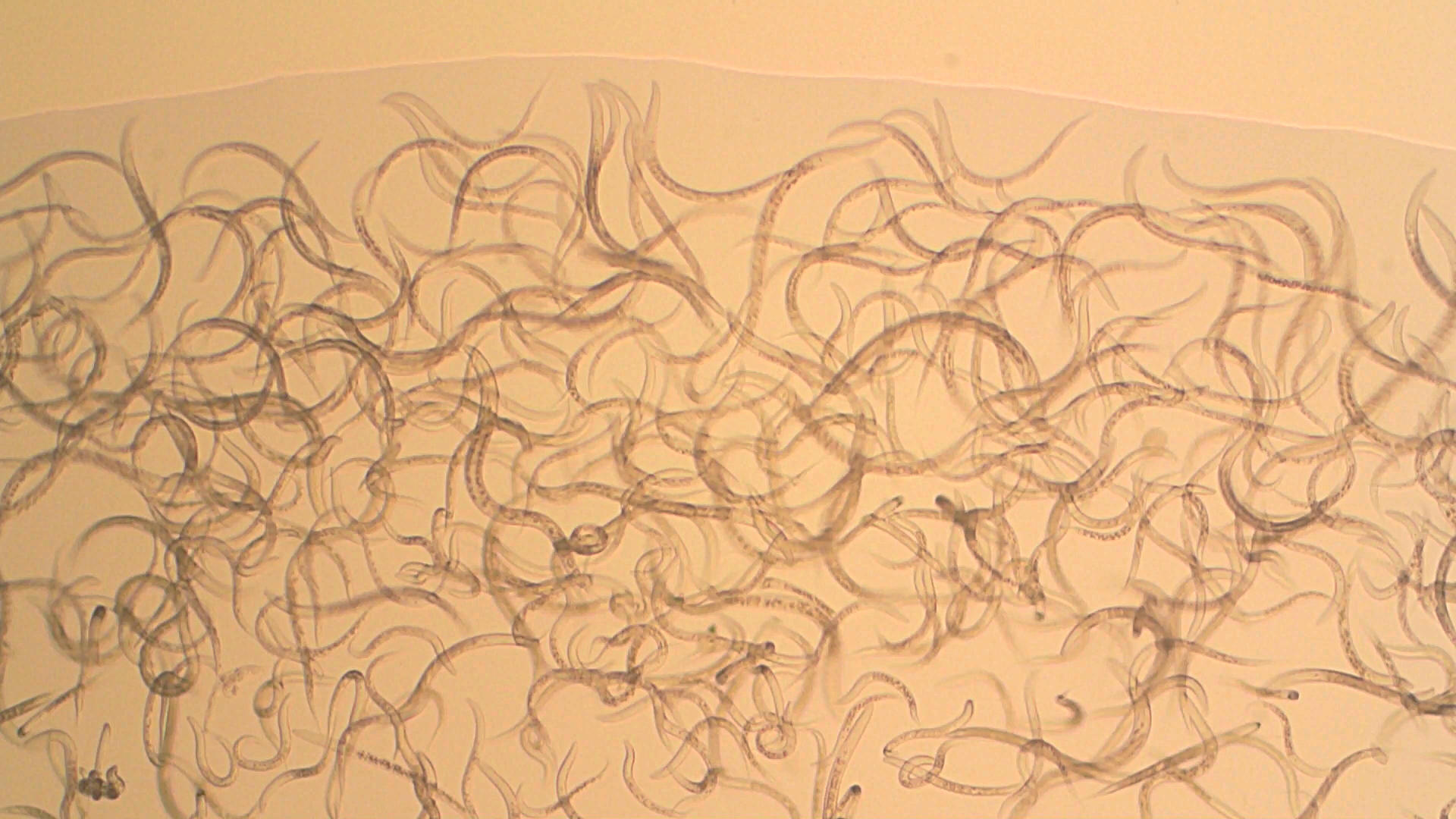}
\par\end{centering}
\caption{\emph{\label{fig:C.-elegans}C. elegans }does exhibit bordertaxis
and individual nematodes oscillate at the border. However, no wave
is observed.}

\end{figure}

\section{Supplemental movies description}

The movies are available at \citep{PESHKOV:2021:ID1057}. All movies
are in real time, except movie 6.
\begin{description}
\item [{Movies}] 1-3: 250 \textgreek{m}l droplet at different moments of
evaporation. Initial density of nematodes $d=30.7\pm3.3\mathrm{n/\mu l}$.
\end{description}
\begin{enumerate}
\item Droplet at time t=1 min. The motion of the nematodes is totally random.
Corresponds to figure 1.a) in the main text.
\item Droplet at time t=20 min. Percolation of the metachronal wave. Corresponds
to figure 1.b) in the main text.
\item Droplet at time t=60 min. A fully developed metachronal wave move
along all the border. Corresponds to figure 1.c) in the main text.
\end{enumerate}
\begin{description}
\item [{Movie}] 4: Metachronal wave observed under a microscope with 4x
magnification. Corresponds to figure 1.d) in the main text.
\item [{Movie}] 5: A shallow droplet with a formed metachronal wave is
observed from the side. The deformation of the surface of the droplet
can be seen.
\item [{Movie}] 6: Rotation of fluorescent particles in the central part
of the droplet from the flow formed by the nematodes participating
in the metachronal wave. The video is accelerated ten times.
\item [{Movie}] 7: Two \textquotedbl defects\textquotedbl{} in the metachronal
wave can be observed which lead to the formation of two counter-rotating
metachronal waves. A 750 \textgreek{m}l droplet at initial density
of nematodes $d=15$ n/\textgreek{m}l. 
\item [{Movie}] 8: Two counter-rotating waves in consecutive diameters.
A 250 \textgreek{m}l droplet at initial density of nematodes $d=31$
n/\textgreek{m}l.
\item [{Movie}] 9: \emph{C. elegans} observed under a microscope at the
border of a droplet. Individual nematodes oscillate at the border,
but no synchronization was ever detected.
\end{description}
\bibliographystyle{apsrev4-2}
\bibliography{anton_references}